\documentclass[
reprint,
amsmath,amssymb,
aps,
prb,
]{revtex4-2}

\usepackage{graphicx}
\usepackage{dcolumn}
\usepackage{xfrac}
\usepackage{bm}

\begin{document}
	
	\preprint{APS/123-QED}
	
	\title{The Structural Behavior of Physisorbed Metallenes}
	
	\author{Pekka Koskinen}
	\email{pekka.j.koskinen@jyu.fi}
	\affiliation{Nanoscience Center, Department of Physics, University of Jyv\"askyl\"a, Finland.}
	\author{Kameyab Raza Abidi}
	\affiliation{Nanoscience Center, Department of Physics, University of Jyv\"askyl\"a, Finland.}
	
	\date{\today}
	
	\begin{abstract}
Atomically thin metallenes have properties attractive for applications, but they are intrinsically unstable and require delicate stabilization in pores or other nano-constrictions. Substrates provide solid support, but metallenes' wanted properties can only be retained in weak physisorption. Here, we study the $45$ physisorbed, atomically thin metallene structures in flat and buckled lattices using a sequential multi-scale model based on density-functional theory calculations. The lattices are mostly buckled but flat for a handful of elements such as Na, K, Rb, Ag, Au, and Cd, depending on physisorption strength. Moreover, under certain conditions, the structure can be controlled by applying biaxial tensile stress parallel or an electric field normal to the surface. The stress reduces the threshold of adhesion strength required to flatten a buckled lattice, and the electric field can be used to increase that threshold controllably. Our results help provide fundamental information about the structures of physisorbed metallenes and suggest means to control them at will by suitable substrate choice or tuning of experimental parameters.
	\end{abstract}
	
	\maketitle

Metallenes are atomically thin, two-dimensional (2D) layers of metals with alluring properties for electronic, catalytic, biomedical, and plasmonic applications.\cite{liu2020metallenes,ta2021insitu,liu2022novel} Unlike covalent 2D materials like graphene or transition metal dichalcogenides \cite{castro2009electronic,manzeli20172d}, their non-directional metallic bonding and lack of layered bulk structure make them tricky to synthesize and stabilize.\cite{koskinen2015plenty} Yet their synthesis has been achieved by etching, 2D growth, and electron irradiation of alloys.\cite{wang2019free,antikainen2017growth,zagler2020cuau,wang20202D,nevalaita2020free,kashiwaya2024synthesis} Stabilization approaches have included graphene pores and other constrictions \cite{zhao2014free,liu2022novel,mendes2024situ}---and of course substrates.\cite{sharma2022synthesis} 

To retain the 2D metallene properties, substrates must provide support without affecting the metallene electronic structure.\cite{schultz2013chemical} This requirement calls for physisorption. With weak binding energies (some tens of meV/\AA$^2$) and large adsorption heights ($3-4$~\AA), physisorption is governed by van der Waals (vdW) forces and lacks chemical bonding. Examples of substrates physisorbing many molecules include metal oxides, zeolites, metal-organic frameworks, and many carbon-based materials.\cite{inert_substrate_1, inert_substrate_2,dresselhaus1999hydrogen,davis2014zeolites, kitagawa2014metal} Still, van der Waals forces can mechanically influence the supported material, modify its structure, and thereby change electronic properties.\cite{korhonen2016limits,zhan2022non} Currently, the trends of the structural behavior of physisorbed metallenes remain unknown.

Therefore, in this article, we ask \emph{what are the trends in the structural behavior of 45 physisorbed, atomically thin metallenes, and can they be controlled?} We address this question with a sequential multi-scale model built upon density-functional theory simulations. The results suggest that weak physisorption can flatten buckled lattices for nearly ten metallenes. The flattening can be further controlled by applying tensile strain or an external electric field. The results provide necessary insight into the structural behavior of physisorbed metallenes and offer the understanding to control metallene structure by suitable experimental design.

Usually, the method to address this type of question would be straightforward density-functional theory (DFT).\cite{vanin2010graphene,penev2021theoretical} However, vdW interaction is tricky for DFT, and reliable and transferable results often require beyond-DFT methods.\cite{olsen2011dispersive,tao2018modeling} Dedicated vdW-DFT exchange-correlation functionals provide reasonable results, although sometimes with compromised accuracies.\cite{andersson1998van,langreth2005van} Also, large supercells required to address lattice mismatch make brute-force DFT for systematic simulations spanning much of the periodic table computationally expensive.\cite{nevalaita2018atlas,nevalaita2018beyond,nevalaita2019stability} Finally, systematic calculations are impractical because no single substrate can serve as a universal benchmark to study physisorption for all metallenes. Therefore, to evade these problems, we use a sequential multiscale approach by combining DFT calculations of pristine metallenes with a model substrate. Apart from predictive power, such an approach helps interpret experiments for various substrates and metallenes. 

\begin{figure}[tb!]
    \centering
\includegraphics[width=0.95\linewidth]{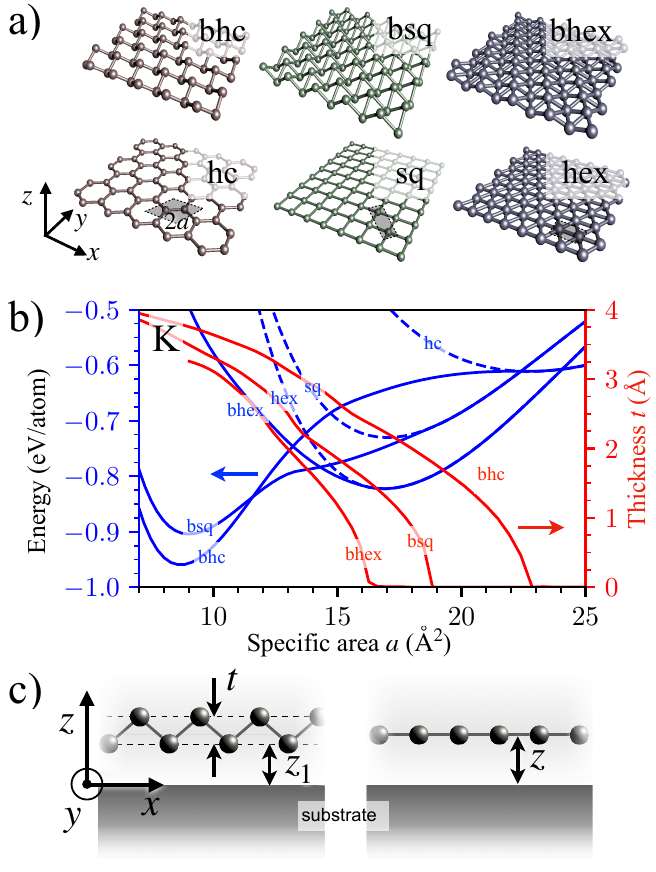}
    \caption{Structures of physisorbed metallenes. a) Six metallene lattices studied here: honeycomb (hc), square (sq), hexagonal (hex), buckled honeycomb (bhc), buckled square (bsq), and buckled hexagonal (bhex). The shaded area shows the two-atom computational cell. The specific area per atom $a$ (half the shared areas) is a free parameter (lattice constant is not fixed). b) The unsupported DFT cohesion energies per atom for the six lattices (left axis) and the thicknesses of buckled lattices (right axis) for K as a function of the specific area. c) The schematics of buckled (left) and flat (right) lattices supported by a model substrate.}
    \label{fig:systems}
\end{figure}

We considered the $45$ metallenes in six crystalline lattices: honeycomb (hc), square (sq), hexagonal (hex), and their buckled counterparts (Fig.~\ref{fig:systems}a). All these $270$ lattices were simulated by the QuantumATK DFT code, using PBE exchange-correlation functional, PseudoDojo pseudopotentials with LCAO basis set, and energy convergence criterion of $10^{-8}$~eV.\cite{perdew1996generalized, smidstrup2019quantumatk} Ref.~\cite{gentle} used similar parameters to investigate free-standing metallenes' dynamical stabilities via phonon spectra, but the physisorption scrambles the phonon dispersions; for this reason, here we consider only static energies. The $k$-point sampling was $13\times 13\times 1$ with a two-atom periodic cell of $20$~\AA\ length in the vertical direction and area $2a$ in the lateral direction (Fig.~\ref{fig:systems}a), adopting the lattice constants of Ref.~\cite{gentle}. All structures were optimized to forces $<10^{-6}$~eV/\AA\ by the BFGS algorithm.\cite{liu1989limited} Calculations using these parameters suffice well for our subsequent multiscale modeling purposes.\cite{abidi2022optimizing} Ultimately, these calculations provided the numerical expressions for the lattice energies $E^L(a)$ and the buckling thicknesses $t^L(a)$ for all $45$ elements and $L\in \{\text{hc, sq, hex, bhc, bsq, bhex}\}$ (Fig.~\ref{fig:systems}b).

The ground states of the unsupported lattices were mostly bhc, except for Ti, Zr, V, Nb, and Fe, where they were bsq. The flat ground state was nearly always hex, with $3$ to $40$~\%\ smaller cohesion energies. The energy differences between the flat and the buckled ground state lattices were roughly proportional to 3D bulk cohesion. These differences could often be argued by changes in coordination numbers, but not always. We did not determine lattice constants separately; the energies $E^L(a)$ were forwarded directly to subsequent multiscale modeling.

For the substrate model, we adopted the Lennard-Jones potential for pairwise atomic interactions between the metallene and the substrate.\cite{lennard1932processes} We integrated the potential over a homogeneous substrate and metallene layers and obtained the adsorbate energy as $V(z)=5V_1/3\times [2/5\times (\sigma/z)^{10}-(\sigma/z)^4]$.\cite{koskinen2014graphene} Here $\sigma$ governs the interaction length scale and $V_1$ is the adhesion strength, which we will adopt as the main parameter characterizing the substrate-metallene interaction.  

Upon assuming that adhesion is independent of adsorbate density, the total energy per atom of the adsorbed metallene becomes 
\begin{equation}
E_{tot}^L(a)=E^L(a)+E_{adh}(t^L(a)).
\label{eq:model}
\end{equation}
Here 
\begin{equation}
    E_{adh}(t)=\min_{z_1} \left\{ \frac{1}{2}[V(z_1)+V(z_1+t)] \right\}
    \label{eq:Eadh}
\end{equation}
is the mean atomic binding energy of a lattice with thickness $t$ (Fig.~\ref{fig:systems}c). As Eq.~(\ref{eq:model}) suggests, adhesion affects $t$ through energy optimization with respect to $a$. 

We validated this model against DFT calculations using the Grimme DFT-D3 functional.\cite{GrimmeDFTD3} To this end, we optimized physisorbed hex and bhc lattices of Au and K to forces $0.05$~eV/\AA\, using graphene as a prototypical substrate.\cite{yin2009bright,lin2010characterizing} The validation systems were $\text{C}_{32}\text{Au}_{12}$ for Au(hex) ($4.4$~\%), $\text{C}_{32}\text{Au}_{24}$ for Au(bhc) ($3.4$~\%), $\text{C}_{12}\text{K}_4$ for K(hex) ($-3.3$~\%), and $\text{C}_{12}\text{K}_4$ for K(bhc) ($-3.8$~\%); values in the brackets were the strains in the metallene. As expected, the validation systems were prototypically physisorbed: the metal atoms reside well over $3$~\AA\ above the substrate, and the adhesion energies are around $40\ldots 50$~meV/\AA$^2$.\cite{amft2011adsorption,settem2024gold} Such physisorption is weak enough to leave the metallene geometric and electronic structures intact, enabling us to benefit from metallenes' unique properties. Fitting vdW-DFT to the model of Eq.~(\ref{eq:model}) gave parameters $V_1=0.28$~eV, $\sigma=3.36$~\AA\ for Au and $V_1=0.70$~eV, $\sigma=3.12$~\AA\ for K. The structures optimized by vdW-DFT and the multiscale model agreed well: the mean absolute errors were only $0.4$~meV/\AA$^2$ for adhesion energy and $0.03$~\AA\ for atomic positions. 


Despite the monolayer thickness, graphene was a reasonable substrate for model validation. We repeated the calculation using two- and three-layer graphite, but adding layers increased adhesion less than $2$~meV/\AA$^2$. Translating metal atoms from the top to the bridge and the hollow sites affected the adhesion less than $1$~meV/\AA$^2$. Also strains of $-10$~\%\ ($\text{C}_4\text{Au}_4$) and $3.4$~\%\ ($\text{C}_{32}\text{Au}_{24}$) for Au(hex) affected adhesion less than $3$~meV/\AA$^2$. Supported by literature, such small energy corrugations vindicate the translational invariance of the model substrate.\cite{ouyang2021registry} 

Literature suggests that, despite being challenging to model by an ab initio approach, the van der Waals interaction can be successfully described by potentials of simple functional form.\cite{andersson1998van,grimme2004accurate,grimme2006semiempirical,GrimmeDFTD3} Moreover, the van der Waals adhesion energy per unit area is surprisingly indifferent to the details of atomic structures.\cite{tang2022high} Such notions imply that---with appropriate parameters for a given substrate and metallene pairs---the model is valid to describe the energetic and geometric properties of physisorbed metallenes. For chemisorption, as mentioned, the model becomes invalid.


\begin{figure}[t!]
    \centering
\includegraphics[width=0.9\linewidth]{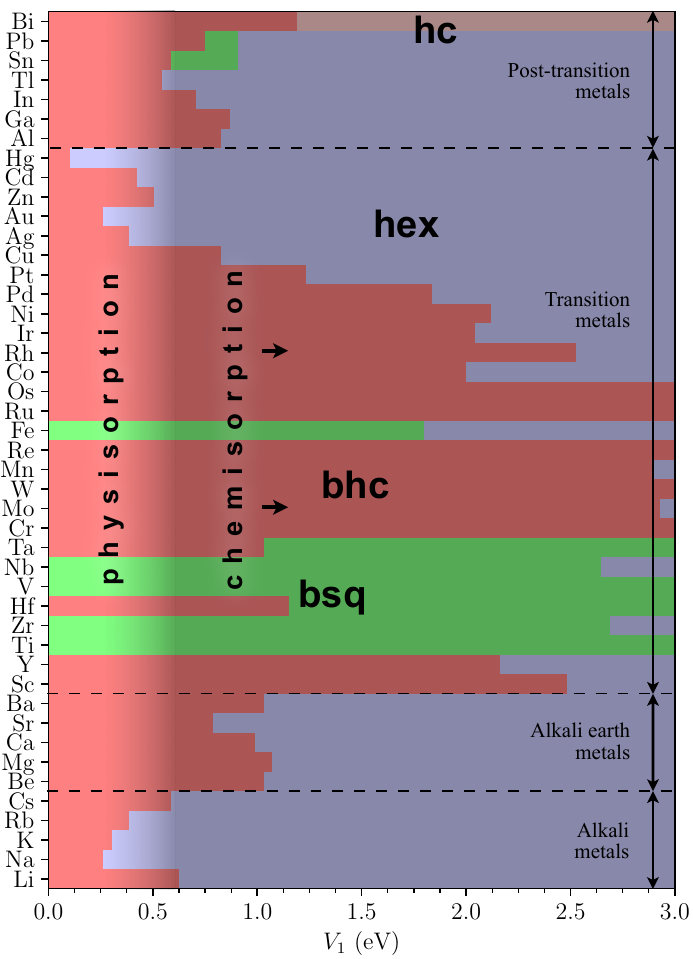}
    \caption{The phase diagram of physisorbed metallenes, showing the lowest-energy lattices at given adhesion strength $V_1$. For completeness, $V_1$ is shown up to values where most lattices become flat.}
    \label{fig:phase_diagram}
\end{figure}

We then applied the model to all $45$ metals and six lattices. The substrate was characterized by the adsorption strength parameter $V_1$. As bond lengths and strengths in chemical bonding can change continuously, there is no precise threshold at which physisorption turns into chemisorption.\cite{gong2010first} For the sake of simplifying the discussion, we here set the threshold at $V_1=0.5$~eV.\cite{gomer1975approaches} The physisorption heights are usually $z\gtrsim 3$~\AA\ and they vary only slightly. In what follows, we fixed $\sigma=3.2$~\AA\, representing a typical adhesion distance; varying $\sigma$ in the range $2-4$~\AA\ affected the results only nominally.

The model enabled optimizing all 270 metallenes systematically and constructing a phase diagram for the ground state lattices as a function of $V_1$ (Fig.~\ref{fig:phase_diagram}). We display the phase diagram for $V_1=0\ldots 3$~eV to convey a complete picture of the structural trends. At the weak adsorption limit, buckled honeycomb is the ground state for most elements, except for the buckled square for Ti, Zr, V, Nb, and Fe. Structural changes under physisorption remain small for all metallenes except for Na, K, Rb, Ag, Au, Cd, and Hg, which can flatten at reasonably small values of $V_1$. Due to its known challenges in DFT, we consider Hg cautiously and will omit its further analysis.\cite{gaston2006lattice} Elements in the early and middle transition metal series can be flattened only by strong chemisorption, which doesn't fall within the scope of this article.

\begin{figure}[t!]
    \centering
\includegraphics[width=\linewidth]{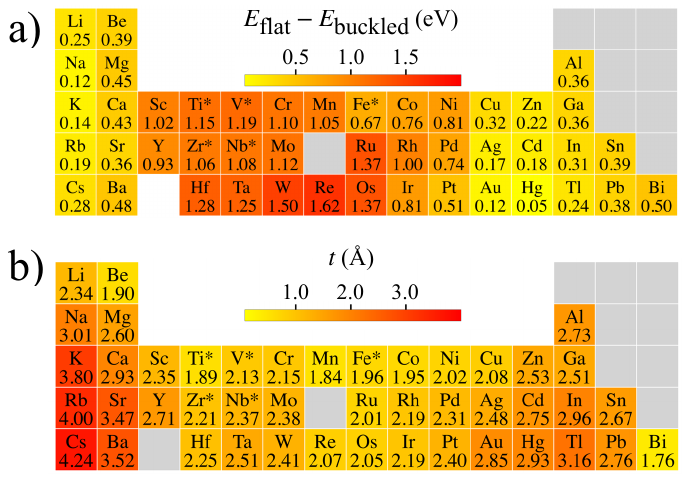}
    \caption{Energetic and structural trends. a) Heatmap of the energy differences between the flat and the buckled ground state lattices for the investigated $45$ elements. The buckled ground state for the elements with an asterisk is bsq; for others, it is bhc. b) The thicknesses $t$ of the lowest-energy buckled lattices.}
    \label{fig:energies_thickness}
\end{figure}

The simple rationale for the phase diagram is that the substrate interaction favors flat lattices by lowering their energy compared to buckled lattices. {Approximately, a lattice flattens when the energy difference between unsupported flat and buckled structures vanishes. The difference equals 
\begin{equation}
E_{adh}(0)-E_{adh}(t)\approx V_1-\frac{1}{2}[V(\sigma)+V(\sigma+t)] \approx 0.4\times V_1  
\end{equation}
for the typical values of $t/\sigma\approx 0.6$ (Fig.~\ref{fig:energies_thickness}b), leading to an estimate for the flattening criterion as
\begin{equation}
    V_1 \gtrsim (E^\text{flat}-E^\text{buckled})/0.4.
    \label{eq:rule-of-thumb}
\end{equation}}
This estimate, which can be confirmed by juxtaposing Figs. \ref{fig:phase_diagram} and \ref{fig:energies_thickness}a, provides a particularly useful rule-of-thumb estimate for quick reference. The buckling thicknesses themselves reside between $t\approx 2\ldots 4$~\AA\ (Fig.~\ref{fig:energies_thickness}b). Such thickness differences between flat and buckled metallenes are well distinguishable by experimental scanning probe techniques.

\begin{figure}[t!]
    \centering
    \includegraphics[width=0.9\linewidth]{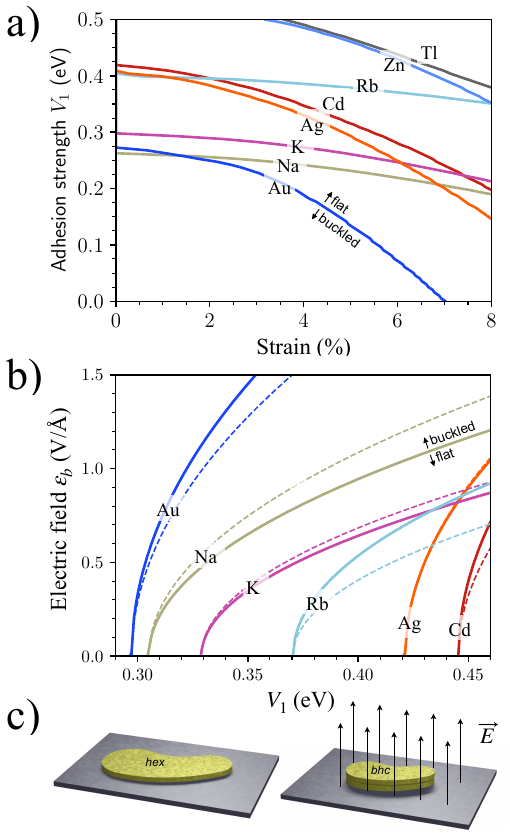}
    \caption{Controlling the structure of physisorbed metallenes. a) The minimum adhesion strength required to flatten a buckled lattice at a given {biaxial lateral strain}. b) The electric field $\varepsilon_b$ required to buckle a flat lattice at given $V_1$ for DFT (solid lines). Comparison is done for the model of Eq.~(\ref{eq:epsb}) calculated with $\alpha_{hex}$ from Ref.~\cite{nevalaita2018beyond} and $\chi=1.25$ (dashed lines). c) Electric field -induced buckling of a finite-size physisorbed patches from hex to bhc reduces lateral area almost by $50$~\%.}
    \label{fig:modify}
\end{figure}

Moreover, the substrate alone does not govern the metallene structure; it can also be controlled.

First, the structure can be controlled by applying {biaxial tensile strain}. Strain can be applied by external confinement \cite{zhao2014free,nevalaita2019stability} or by corrugated potential energy landscape \cite{yin2015simple}, as evidenced by ubiquitous moir\'e patterns in 2D heterostructures.\cite{carr2020electronic} Mechanical bending can be used to control strain even \emph{in situ}.\cite{wang2013situ} {To investigate the effect of strain, we compared the energy differences between the unstrained flat lattices and the biaxially strained buckled lattices upon physisorption.} It turned out that Na, K, Rb, Ag, Au, and Cd were flattened at even weaker physisorption than at zero strain (Fig.~\ref{fig:modify}a). {Being simple metals, Na, K, and Rb with their jellium-like electronic structure are relatively insensitive to geometric details. 
Therefore, the flattening threshold is unresponsive to the geometric changes due to strain. In contrast, being late transition metals with more directional d-orbital bonding, Au, Ag, and Cd are more sensitive to strain.} The strain has a limited effect on flattening, but additional metals Zn and Tl appear in the physisorption window for strains $\gtrsim 5$~\%. 

Second, the structure can be controlled by an electric {field normal to the surface, along the [001] direction}. The energy in the electric field $\varepsilon$ changes like $E^L_{tot}(a,\varepsilon)=E^L_{tot}(a)-\tfrac{1}{2}\alpha \varepsilon^2$, where $\alpha$ is the vertical polarizability of the metallene, which is slightly larger for the buckled lattice with its more responsive electron density between the layers. Therefore, applying an electric field may switch the ground state from flat to buckled.

We demonstrated this scenario by calculating  
$E^L(a,\varepsilon)$ under different electric fields for Au, Na, K, Rb, Ag, and Cd by DFT. The DFT energies $E^L(a,\varepsilon)$ were inserted into the model of Eq.~(\ref{eq:model}) and solved for the smallest electric field $\varepsilon_b$ that satisfied the buckling condition $\min_a [E^\text{bhc}_{tot}(a,\varepsilon_b)]\leq \min_{a'} [E^\text{hex}_{tot}(a',\varepsilon_b)]$ for given $V_1$. As a result, several elements within the physisorption window allow structural control at sensible electric fields $\varepsilon_b$ (Fig.~\ref{fig:modify}b).\cite{weintrub2022generating} In particular, control over buckling implies simultaneous control over the lateral area: buckling decreases the area almost by $50$~\%. Such control signifies tuning the size of metallene patches by turning a knob (Fig.~\ref{fig:modify}c), which is useful for applications based on plasmons, electronics, and structural control.\cite{greybush2017plasmon,murthy2020direct,mirigliano2021electrical}

The computed trends on $\varepsilon_b$ can be understood analytically. It is straightforward to derive an expression for $\varepsilon_b$ as 
\begin{equation}
\varepsilon_b=\sqrt{\frac{2\Delta E}{\alpha_{hex}(\chi-1)}}.
\label{eq:epsb}
\end{equation} 
Here $\Delta E$ is the energy difference between the physisorbed buckled and flat lattices and $\chi=\alpha_{bhc}/\alpha_{hex}$ is the polarizability ratio, where $\alpha_{bhc}$ and $\alpha_{hex}$ are the polarizabilities of bhc and hex lattices. As discussed in Ref.~\cite{nevalaita2018beyond}, $\alpha_{hex}$ can be described by a dipole interaction model, suggesting an expression
\begin{equation}
\alpha_{hex}=\frac{d^3}{4S}\sqrt{1+\frac{8S\alpha_\text{free}}{d^3}}-1 ,
\label{eq:alpha}
\end{equation}
where $\alpha_\text{free}$ is atomic polarizability, $d$ is the bond length of 3D bulk, and $S$ is the lattice sum discussed in Ref.~\cite{monkhorst1976special}. It turned out that, although working well for flat metallenes, the dipole interaction model did not correctly describe thick metallenes and could not directly determine $\chi$, which had to be adopted as a fitting parameter. Using $\alpha_{hex}$ from Ref.~\cite{nevalaita2018beyond} gives the fit $\chi=1.25$, which results in a rough agreement with the DFT results (Fig.~\ref{fig:modify}b). By using a previously fitted trend $\alpha_{hex}=5.34\times d^3$ meV/V$^2$\AA~\cite{nevalaita2018beyond}, we get an approximate but concise expression for the critical field as
\begin{equation}
\varepsilon_b=39.0\times \sqrt{\frac{\Delta E}{d^3} \frac{\text{V}^2{\mathring {\mathrm A}}}{\text{eV}} }.
\label{eq:epsb2}
\end{equation}

To conclude, we investigated physisorbed metallenes using a multiscale model based on energies and structures from DFT calculations. It turned out that the structural behavior of physisorbed metallenes depends on the substrate but can also be controlled. The results indicated ground states are usually buckled, but they can also get flattened for some ten elements if the adhesion is strong enough. Under certain conditions, tensile strain and external electric field can control the metallene structure; this control also means authority over metallene properties and function. Structures bordering the flattening threshold are particularly attractive because their flattening (and buckling) transitions could be triggered by weak external perturbations.

\section*{Conflicts of interest}
There are no conflicts to declare.

\section*{Data availability statement}

Data for this article, including the lattice energies calculated by QuantumATK (https://www.synopsys.com/manufacturing/quantumatk.html), are available via gitlab at https://gitlab.jyu.fi/ldnmm/2dmetals/-/tree/master/2025\%20Metallene\%20Physisorption.

\section*{Acknowledgments}
We acknowledge the Vilho, Yrjö, and Kalle Väisälä Foundation of the Finnish Academy of Science and Letters and the Jane and Aatos Erkko Foundation for funding (project EcoMet) and the Finnish Grid and Cloud Infrastructure (FGCI) and CSC - IT Center for Science for computational resources.

\renewcommand\refname{References}


\providecommand*{\mcitethebibliography}{\thebibliography}
\csname @ifundefined\endcsname{endmcitethebibliography}
{\let\endmcitethebibliography\endthebibliography}{}

\end{document}